\documentclass[sigconf]{acmart}

\makeatletter
\def\@ACM@checkaffil{
    \if@ACM@instpresent\else
    \ClassWarningNoLine{\@classname}{No institution present for an affiliation}%
    \fi
    \if@ACM@citypresent\else
    \ClassWarningNoLine{\@classname}{No city present for an affiliation}%
    \fi
    \if@ACM@countrypresent\else
        \ClassWarningNoLine{\@classname}{No country present for an affiliation}%
    \fi
}
\makeatother
\AtBeginDocument{%
  \providecommand\BibTeX{{%
    \normalfont B\kern-0.5em{\scshape i\kern-0.25em b}\kern-0.8em\TeX}}}

\copyrightyear{2024} 
\acmYear{2024} 
\setcopyright{acmlicensed}\acmConference[HRI '24 Companion]{Companion of the 2024 ACM/IEEE International Conference on Human-Robot Interaction}{March 11--14, 2024}{Boulder, CO, USA}
\acmBooktitle{Companion of the 2024 ACM/IEEE International Conference on Human-Robot Interaction (HRI '24 Companion), March 11--14, 2024, Boulder, CO, USA}
\acmDOI{10.1145/3610978.3640610}
\acmISBN{979-8-4007-0323-2/24/03}

\usepackage{subcaption}
\usepackage{graphicx}
\usepackage{booktabs}
\usepackage{hyperref}
\usepackage{multirow}

\begin{document}
\title{DualTake: Predicting Takeovers across Mobilities for Future Personalized Mobility Services}

\author{Zhaobo Zheng}
\affiliation{%
  \institution{Honda Research Institute USA, Inc}}
\email{zhaobo_zheng@honda-ri.com}

\author{Kumar Akash}
\affiliation{%
  \institution{Honda Research Institute USA, Inc}}
\email{kakash@honda-ri.com}

\author{Teruhisa Misu}
\affiliation{%
  \institution{Honda Research Institute USA, Inc}}
\email{tmisu@honda-ri.com}

\renewcommand{\shortauthors}{Zheng, Akash, and Misu}

\begin{abstract}
  A hybrid society is expected to emerge in the near future, with different mobilities interacting together, including cars, micro-mobilities, pedestrians, and robots. People may utilize multiple types of mobilities in their daily lives. As vehicle automation advances, driver modeling flourishes to provide personalized intelligent services. Thus, modeling drivers across mobilities would pave the road for future society mobility-as-a-service, and it is particularly interesting to predict driver behaviors in newer mobilities with traditional mobility data. In this work, we present takeover prediction on a micro-mobility, with car simulation data.The promising model performance demonstrates the feasibility of driver modeling across mobilities, as the first in the field.
\end{abstract}




\received{8 December 2023}
\received[accepted]{10 January 2024}

\maketitle

\section{Introduction}
To contribute to more sustainable and convenient future cities, more transportation modes are needed \cite{richards2012future, avetisyan2022design}. Many types of micro-mobilities are getting popular, such as e-scooters, mopeds, and e-bikes. They are increasingly recognized as a promising urban transport mode, particularly for their advantages in parking, first/last mile convenience, and short-distance travel \cite{abduljabbar2021role}. They also have broader impacts on fuel efficiency and carbon emissions. A recent study has shown that E-scooter has better emission and operation costs than internal combustion cars, electric cars, and even public electrical buses \cite{d2022adoption}. The micro-mobilities are also a good fit for modern and future society because of their health benefits. The transition from car to electric micro-mobility has been found to increase physical activity and prevent fatal accidents, which outweigh the air pollution exposure \cite{lopez2022health}. With these advantages, social acceptance of mixed types of mobilities, as in shared mobilities or an integration of public transportation has emerged as an alternative for private cars and public transportation\cite{oeschger2020micromobility}.

Human-machine interaction is an important aspect of intelligent vehicles \cite{mehrotra2022human}, that relate to vehicle control, and more recently, personalized experiences and driver state understanding \cite{kun2018human, zepf2020driver}. As automated vehicle (AV) and automated driver assistance systems (ADAS) advance, the driver-vehicle interaction is of high importance to safety, driver trust, and acceptance \cite{becker2017literature,mehrotra2023trust,mehrotra2023does}. Besides research on driver fatigue, distraction, and driving styles \cite{abbas2020driver, kashevnik2021driver, zheng2022identification}, driver takeovers in AV remain critical because it is a fundamental interaction related to safety and user trust \cite{akash2018classification, wu2019take}. Takeover prediction through non-invasive in-vehicle sensors can enhance driving safety and improve user experiences. Pakdamanian et al. used eye movement, heart rate (HR), and galvanic skin response (GSR) to predict driver takeovers \cite{pakdamanian2021deeptake}. They built a deep neural network (DNN) to predict takeover intention, time, and quality, and the accuracy is 96\%, 93\%, and 83\%, respectively. Du et al. predicted takeover performance in conditionally automated driving vehicles \cite{du2020predicting}. They conducted the n-back memory task during conditional automated driving, with gaze, physiological, and facial monitoring. Their best-performing model was random forest (RF), and it had an area under the receiver operating characteristic curve (AUC) of 0.69. This research implicates the function development of driver monitoring in intelligent vehicles. With takeover prediction, intelligent vehicles can potentially adjust driving behaviors to avoid takeovers or facilitate smooth control transition. These applications would enhance user trust, comfort, and experience \cite{jin2021modeling}\cite{li2019investigating}. 

Existing research are all limited to cars, on which current AV and ADAS development focused. It is the same case for driver state understanding in a broader term \cite{lohani2019review}. The future hybrid society requires driver-state monitoring of different mobilities. For now, the majority of users may have not experienced micro-mobility. In the year of 2019, there are about 50 million e-scooter rides in the US \cite{DOTreport}. That is less than 0.15 e-scooter trips per capita in that year, compared to 84.1\% of the US population being a licensed car driver. As a result, most users are novel to micro-mobility, but we still need to provide driver monitoring services on newer mobilities, by modeling user behaviors with data only from traditional mobilities.

Therefore, we propose takeover prediction on micro-mobility with behavioral data from cars, through a deep neural network and transfer learning. This work is novel to demonstrate the feasibility of across-mobility driver monitoring, without data needed from the newer mobility. The rest of the paper is organized as follows: section 2 describes the user study and the dataset we used; Section 3 introduces the behavioral differences between the mobilities; Section 4 presents the machine learning modeling of takeover prediction; and we finally conclude the paper in section 5 with discussions.   

\section{User Study}
We conducted a user study for data collection. A total of 48 participants were recruited in this study, with 28 males and 20 females. They aged between 19 to 69, with an age mean of 33.8 years, and a standard deviation of 11.7 years. The study was approved by the Institutional Review Board at San Jose State University.

\subsection{Apparatus and Scenarios}
\begin{figure}
     \centering
     \begin{subfigure}[]{0.20\textwidth}
         \includegraphics[width=\textwidth]{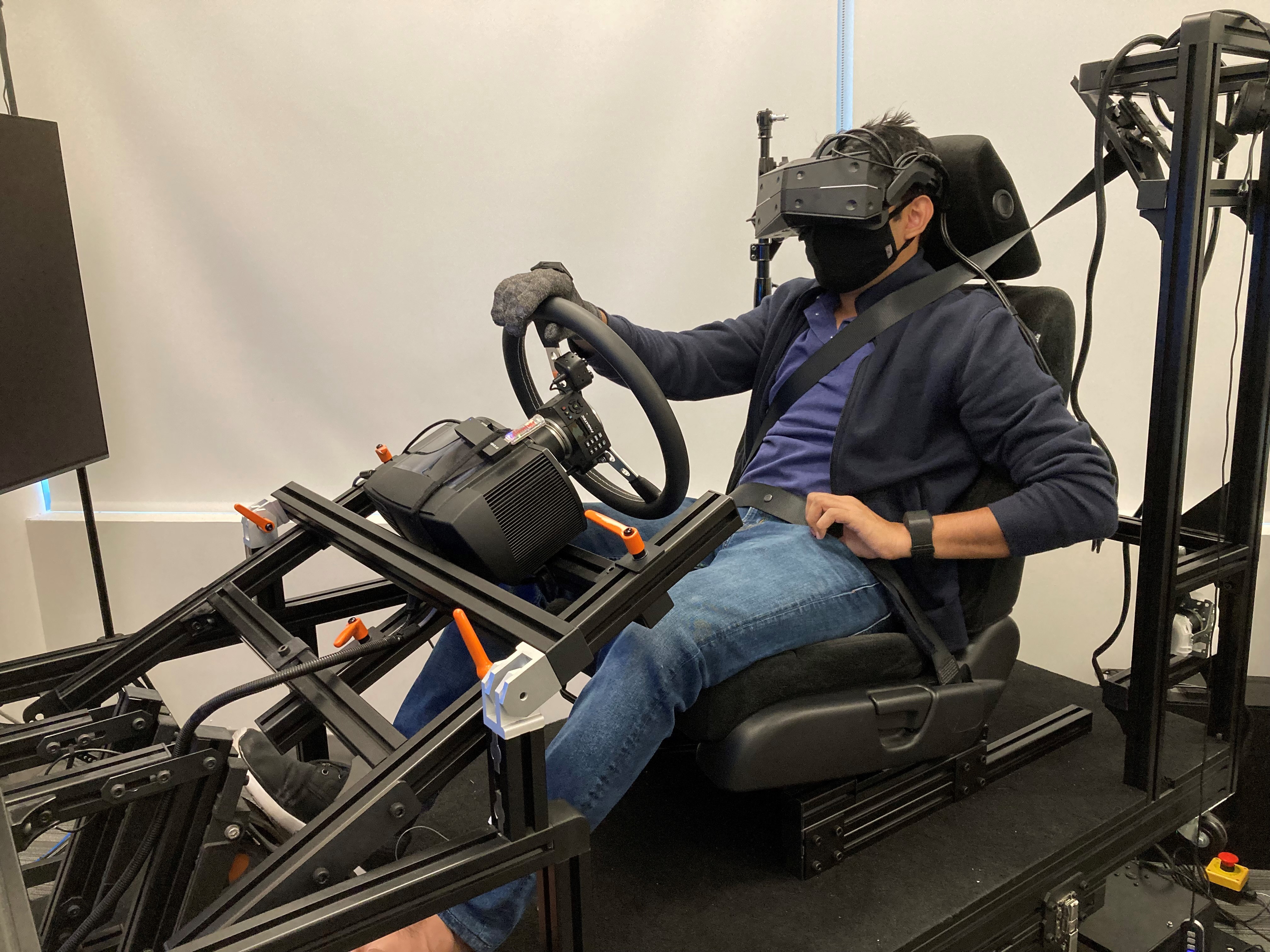}
         \caption{Car Platform}
         \label{car}
     \end{subfigure}\hspace{20px}
     \begin{subfigure}[]{0.20\textwidth}
         \includegraphics[width=\textwidth]{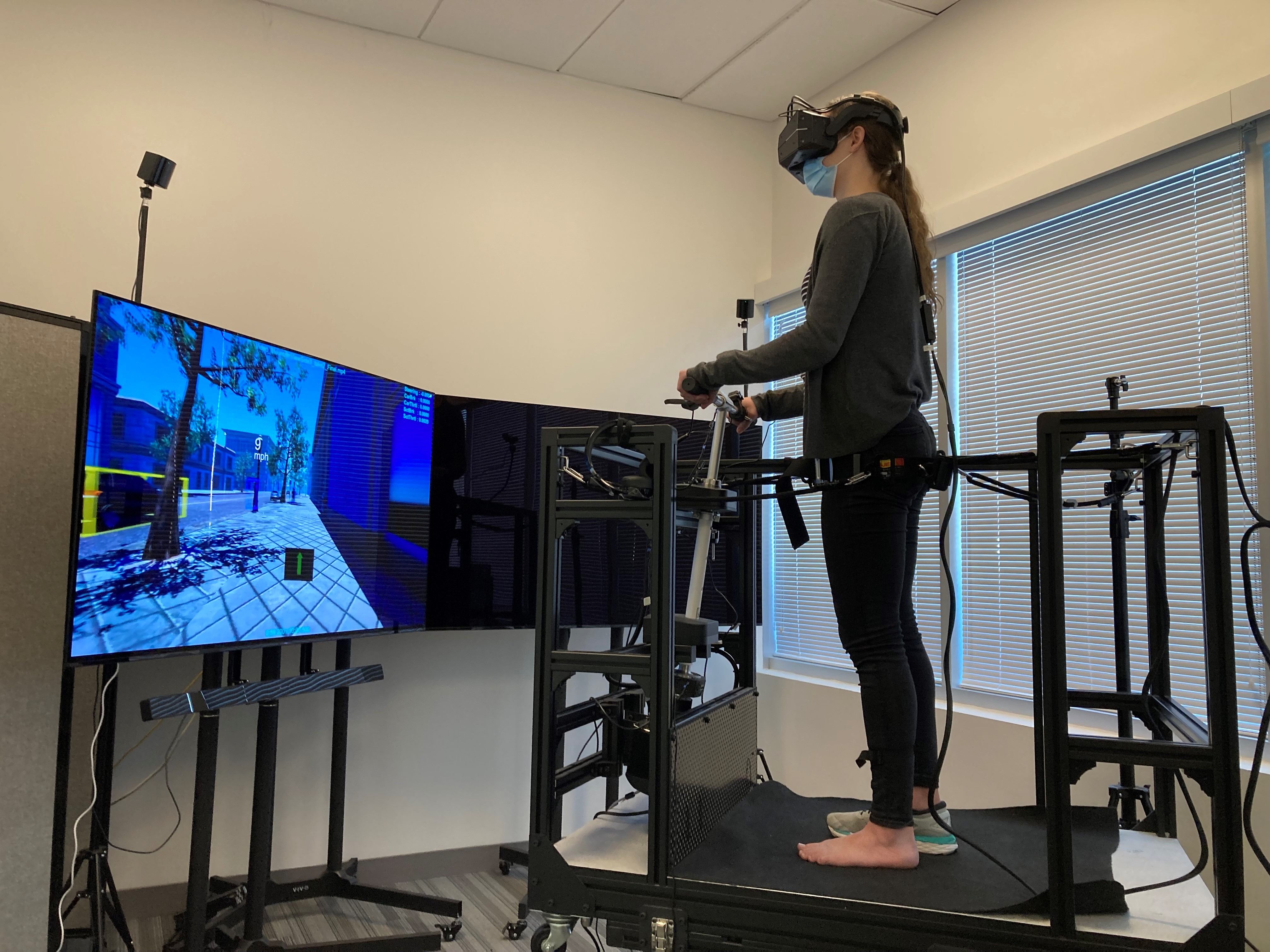}
         \caption{Micro-mobility Platform}
         \label{micromobility}
     \end{subfigure}\vspace{-1em}
        \caption{Motion Base and Mobility Platforms}
        \label{simulator}\vspace{-2em}
\end{figure}

A high-fidelity simulator was integrated for immersive experiences and natural responses. We used a VR headset \cite{StarVROne} to incorporate a wide field-of-view (FoV). A motion base \cite{MotionBase} was used to mimic vehicle motions. The mobility platforms are mounted on the motion base, and the platforms include a steering wheel, throttle, and brake peddles for the car, and steering handle, throttle, and brake buttons for the micro-mobility. The overview of the simulator equipment can be seen in Fig.\ref{simulator}. We integrated a multimodal sensing framework to capture driver behaviors. The sensors include eye gaze tracking, a Shimmer physiological sensor \cite{Shimmer}, and an E4 wristband \cite{E4}. We applied both GSR+ and E4, for better physiological quality on the GSR signals.

\begin{figure}
     \centering
     \begin{subfigure}[]{0.23\textwidth}
         \includegraphics[width=\textwidth]{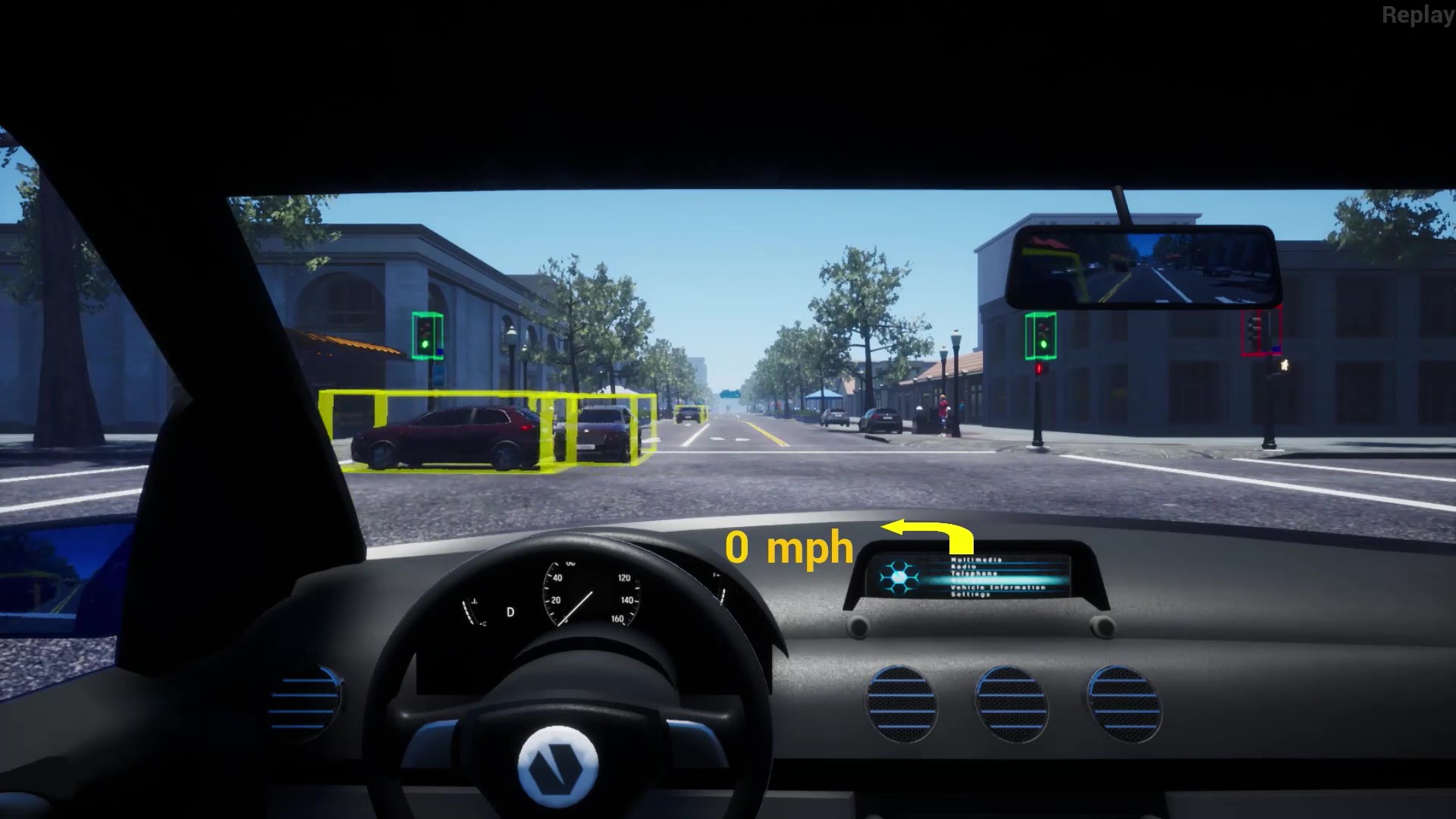}
         \caption{Car Scene}
         \label{carscene}
     \end{subfigure}
     \begin{subfigure}[]{0.23\textwidth}
         \includegraphics[width=\textwidth]{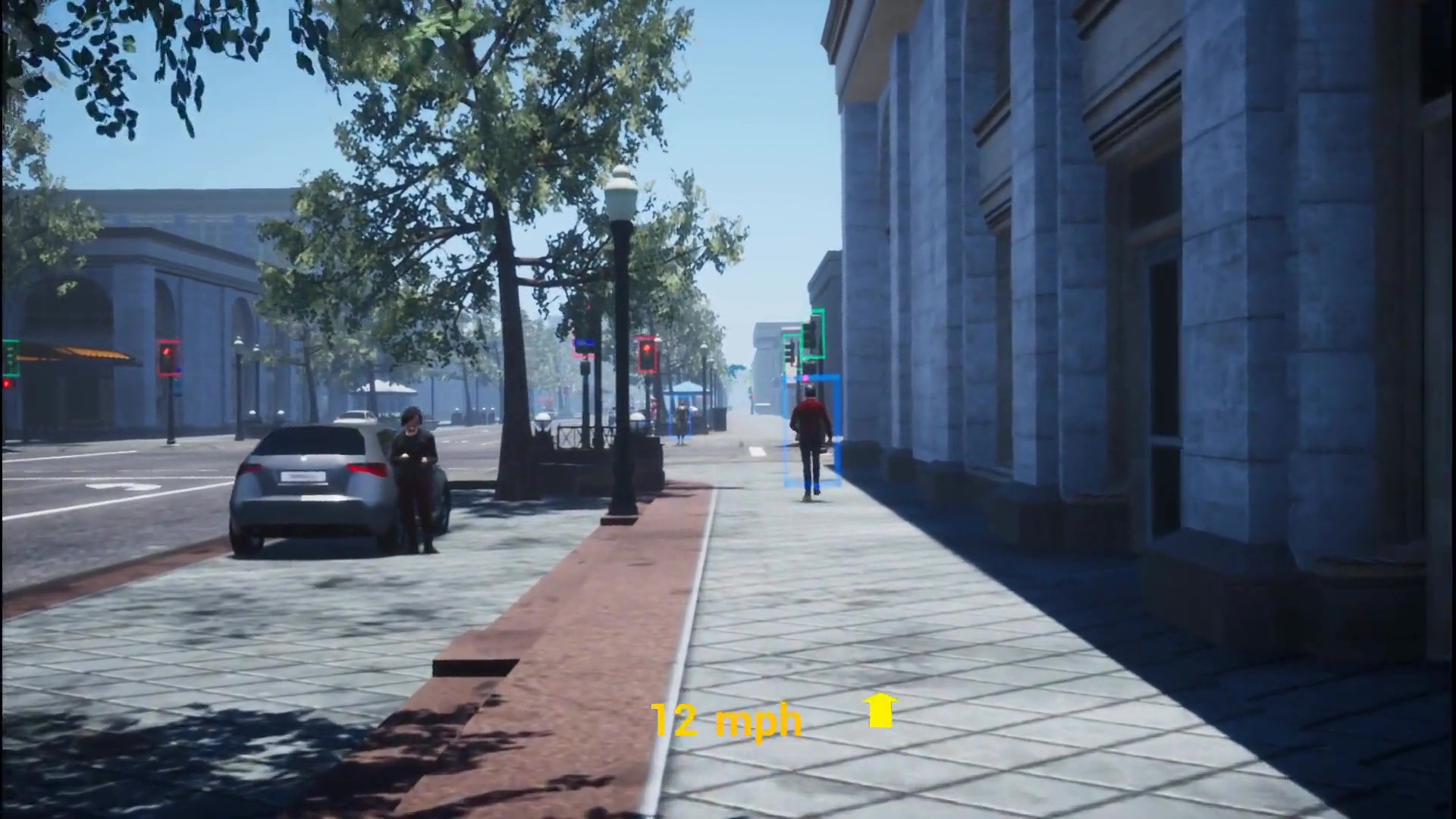}
         \caption{Micro-mobility Scene}
         \label{micromobilityscene}
     \end{subfigure}\vspace{-1em}
        \caption{Example Scenes on Car and Micro-Mobility}
        \label{scenes}\vspace{-1em}
\end{figure}

The participants experienced both the car and the micro-mobility, in a counter-balanced order. The duration of the experiment averaged around 1 hour 50 minutes. We simulated an urban driving environment, with road objects including pedestrians, cars, buildings, roads, sidewalks, trees, traffic lights, traffic signals, and stop signs. The micro-mobility traveled mostly on sidewalks, but it traveled on separate bike lanes too. We designed the study in the urban environment, as the most common user case of micro-mobilities\cite{latinopoulos2021planning}. Some example scenarios for both mobilities are shown in Fig. \ref{scenes}. The participants were asked to monitor the automated drives and takeover when they deemed necessary, with brake or throttle inputs. The automated driving was simulated by the "Wizard of Oz" technique. The AV had driving styles of aggressive and defensive. They differ in driving behaviors such as whether to yield to other road users, and whether to stop at stale green lights. The AV also had a proactive mode, that gave some audio alert to the participant of its intentions, such as "waiting for a pedestrian to pass". The participants experienced the driving styles with and without proactive mode in a counterbalanced order. Driver behavioral response related to driving style and proactive feedback differences are out of the scope of this work. Investigations on those aspects and more technical details can be found in existing research \cite{hunter2023future}.

\subsection{Signal Processing and Feature Extraction}

\begin{table}\footnotesize{
  \caption{Feature List}\vspace{-1em}
  \label{tab:feature}
  \begin{tabular}{ccc}
    \toprule
    Data Modality&Count&Features\\
    \midrule
    GSR&5& $GSR^*$ $SCR_{count}$ \\
    HR&5&$HR^*$ $HRV$\\
    \midrule
    Eye gaze&9&$x_{gaze}^*$ $y_{gaze}^*$ $entropy_{region}$\\
    Gaze Semantics&15&$p_{objects}$ $entropy_{object}$\\
    \midrule
    Maneuver&4&$steering^*$\\
    \midrule
    CAN-Bus&14&$v_{x}^*$ $v_{y}^*$ $\omega_{z}^*$ $Aggressiveness$ $Proactive$\\
  \bottomrule\vspace{-2em}
\end{tabular}}
\end{table}

Of the 48 experiment participants, 42 participants had complete data from all modalities. The missing data are due to sensor malfunctioning or manual operation mistakes, such as physiological sensor recording issues and eye tracking freezes. We then synchronized the signals from different modalities and applied a nearest-neighbor interpolation on short-interval null signals. These null signals are mainly due to blinks. A Z-normalization was performed to eliminate the physical individualized differences.

We extracted a total of 52 features, as shown in Table \ref{tab:feature}. We selected the features based on ablation studies from existing literature \cite{du2020predicting}\cite{zheng2022detection}. An asterisk means that we extracted the mean, standard deviation, minimum, and maximum values as 4 separate features. We used a sliding window of 10 seconds, without window coincidence. The takeover label was defined as whether the participants had throttle or brake inputs in the 3 seconds after the time window. For peripheral physiological signals, we computed the statistical features of the skin conductance level and the heart rate level. We also extracted the skin conductance response episode counts, and heart rate variations, which are highly correlated with human arousal and stress \cite{schaaff2013measuring}. 

\begin{figure}
     \centering
     \begin{subfigure}[]{0.23\textwidth}
         \includegraphics[width=\textwidth]{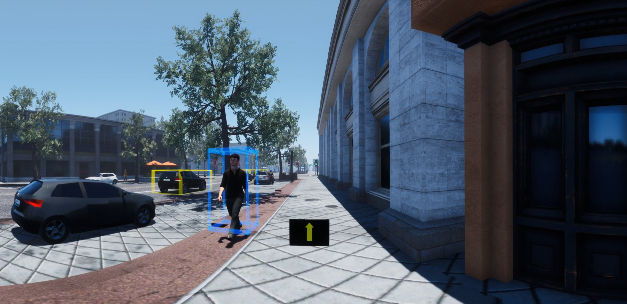}
         \caption{Original VR Frame}
         \label{originalVR}
     \end{subfigure}
     \begin{subfigure}[]{0.21\textwidth}
         \includegraphics[width=\textwidth]{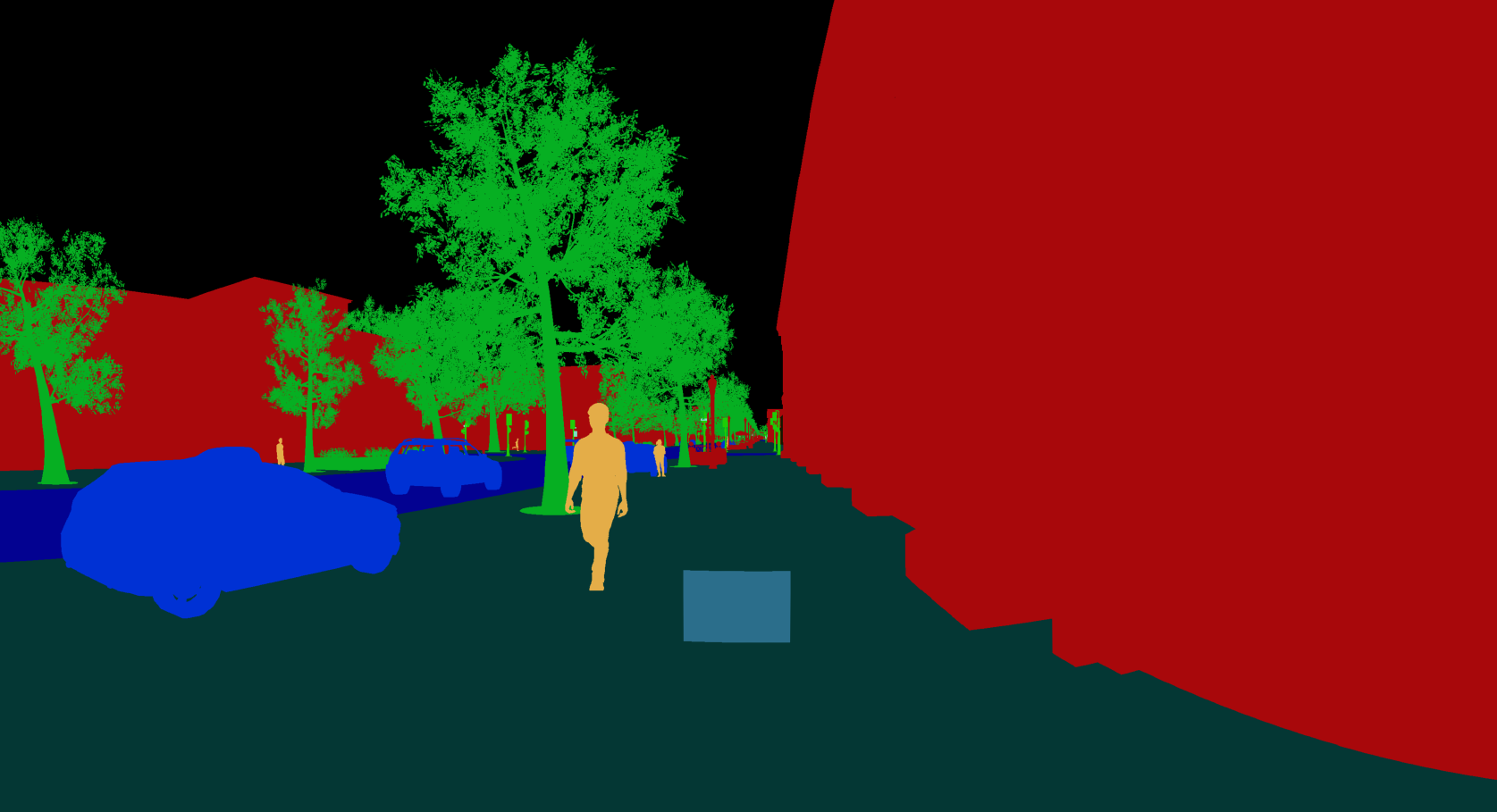}
         \caption{Semantic Segmentation}
         \label{semantics}
     \end{subfigure}\vspace{-1em}
        \caption{Original VR Snapshot and its Semantics Segmentation}
        \label{semantics_comparison}\vspace{-2em}
\end{figure}

The VR headset gaze tracking provides the segment of the scene in the 360 video and the gaze 2d position in that specific segment. We computed the 2D gaze in the 360 video from those signals. We also segmented the 360 frames into 9 different regions, such as top left, and middle right regions. Then we computed the gaze region entropy to estimate the degree of focus of the participant's gaze. Existing research demonstrates that the object people are fixing on is rich in prediction power\cite{koochaki2023learn}. Thus, we extracted the semantic segmentation of each frame, as shown in Fig. \ref{semantics_comparison}. We also computed the gaze entropy for objects. The gaze-related features had a feature extraction window of 1 second because gaze changes much faster than other signals like physiology. We extracted the statistical features of the steering angle inputs from the participants. For CAN-Bus data, we computed the statistical features of the linear velocities on the vehicle and the angular velocity. We also included the aggressiveness of the vehicle, and whether it was proactive.

\section{Statistical Analysis on Driver Behaviors across Mobilities}
\begin{figure*}
     \centering
     \begin{subfigure}[]{0.21\textwidth}
         \includegraphics[width=\textwidth]{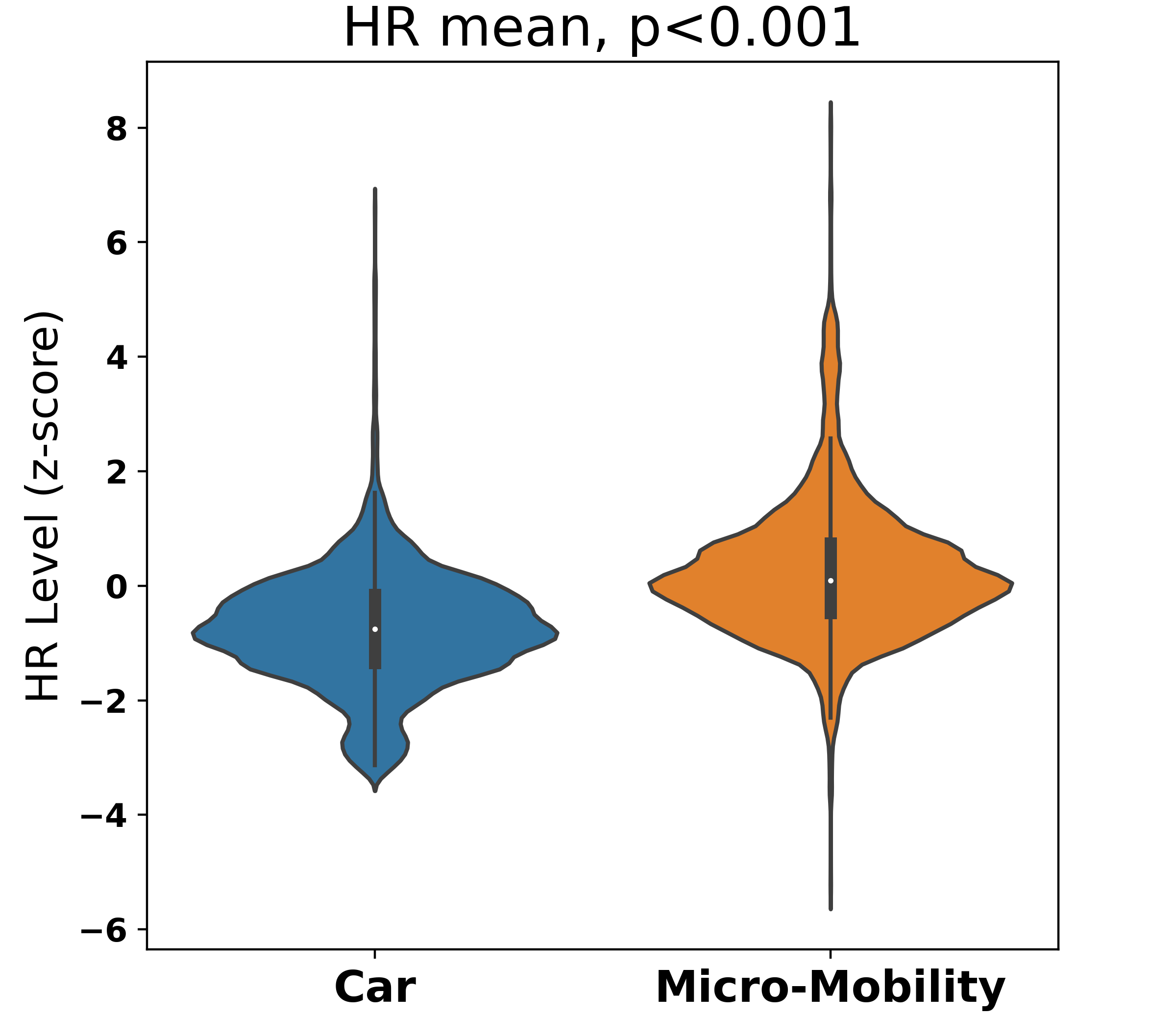}
         \caption{HR mean}
         \label{HR_mean}
     \end{subfigure}
     \begin{subfigure}[]{0.21\textwidth}
         \includegraphics[width=\textwidth]{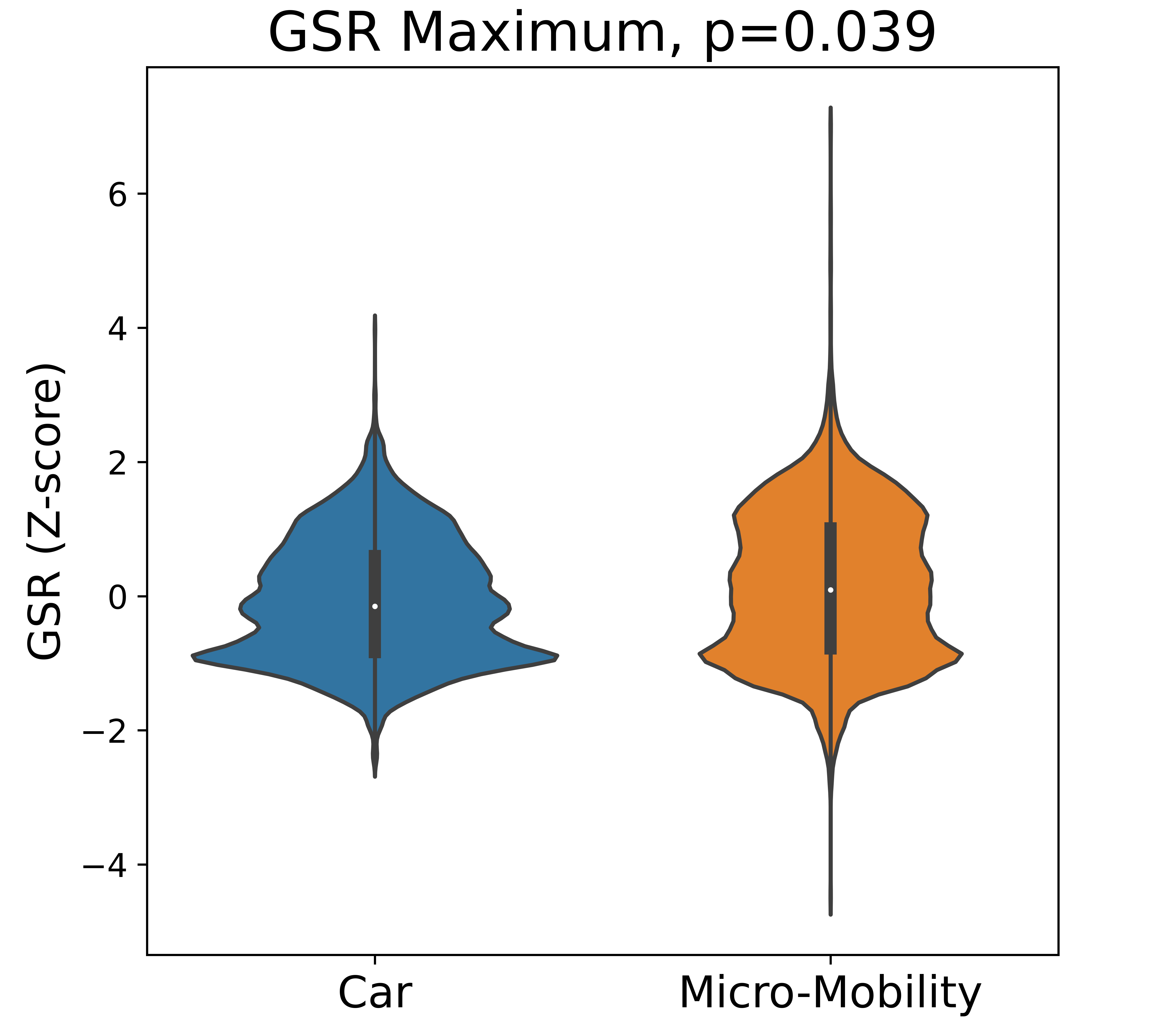}
         \caption{GSR maximum}
         \label{GSR_min}
     \end{subfigure}
     \begin{subfigure}[]{0.21\textwidth}
         \includegraphics[width=\textwidth]{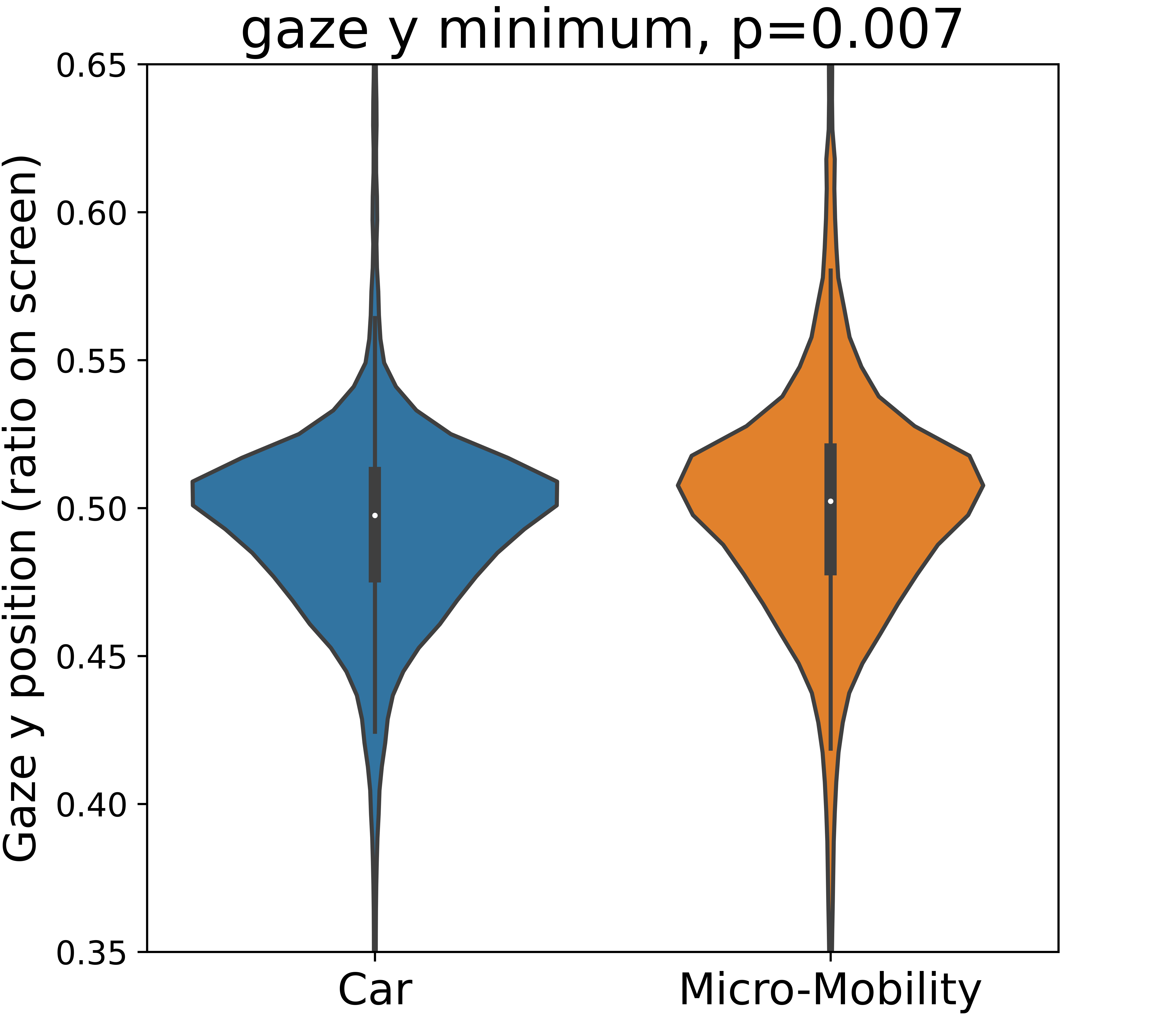}
         \caption{Gaze y min}
         \label{gaze_y_mean}
     \end{subfigure}
     \begin{subfigure}[]{0.21\textwidth}
         \includegraphics[width=\textwidth]{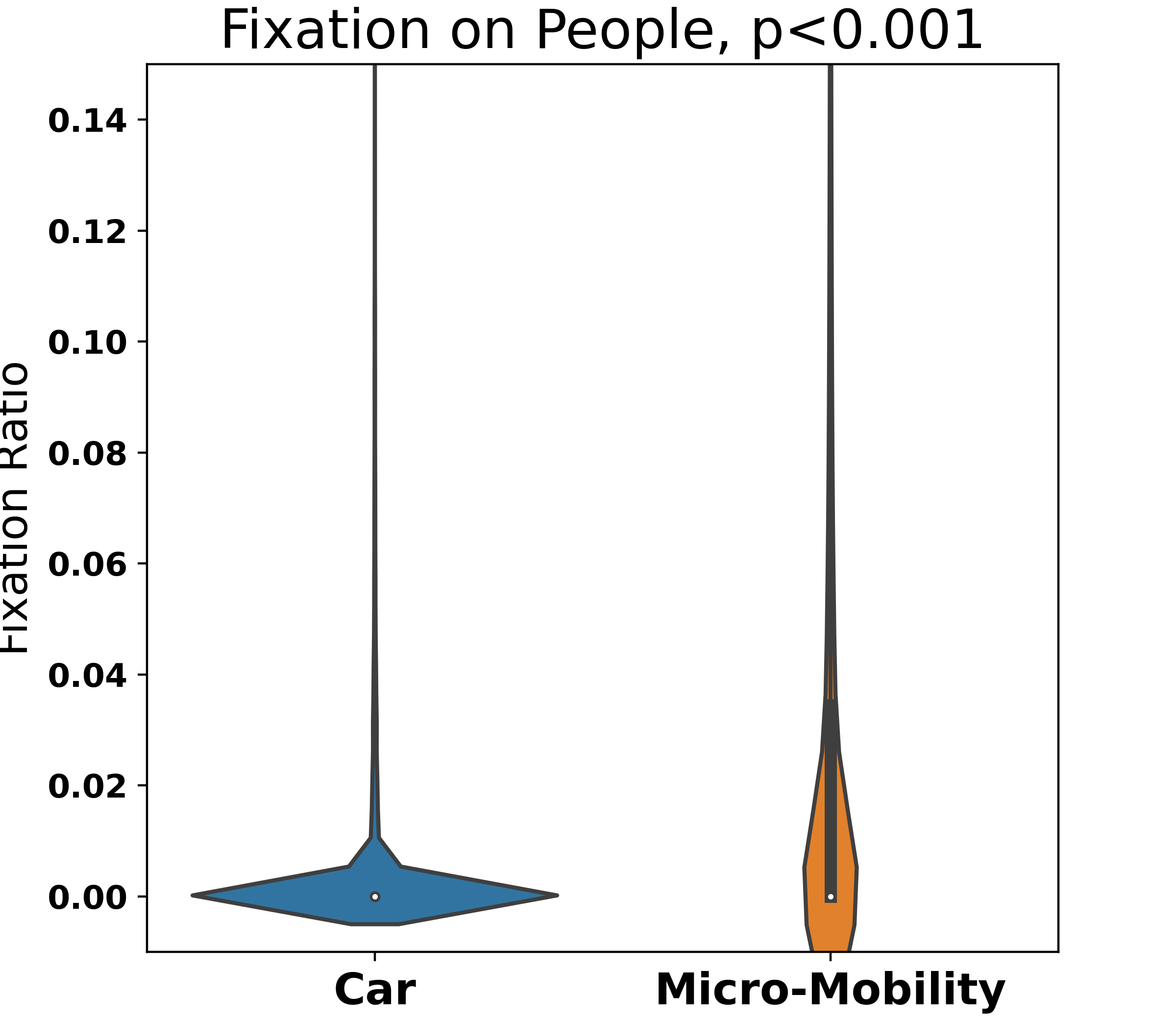}
         \caption{Fixation on People}
         \label{people}
     \end{subfigure}\vspace{-1em}
        \caption{Behavioral Differences Across Mobilities}
        \label{stats}\vspace{-0.5em}
\end{figure*}

To demonstrate different characteristics across the mobilities, we conducted a statistical analysis. We performed a paired t-test on the extracted features and found many statistical significance. We visualize some of the representing results, as shown in Fig. \ref{stats}. First, we observe higher HR level mean (p<0.001) and GSR maximum level (p=0.039) on micro-mobility. In our experiment, the participants sat in the car while they stood in the micro-mobility. The increased physical activity in micro-mobility may have led to this peripheral physiological difference. Then, we found that the minimum gaze position was higher in the micro-mobility (p=0.007). From Fig. \ref{simulator}, the cabin is blocking some parts of the scene while the micro-mobility almost has a free FoV. Participants may have focused more on the lower regions of the scenes because of the cabin. Similarly, very significant differences were found in object fixations. Participants in micro-mobility had way more fixation on people, and they almost did not pay attention to people in cars (p<0.001). Micro-mobility traveled on the sidewalk and there was more interaction and closer distance with other pedestrians. It is worth noticing that significant object fixation differences were also found in fixation on roads (p<0.001), sidewalks (p<0.001), and road signs (p=0.049). The many behavioral differences across mobilities are key factors that make this prediction task challenging. People naturally behave differently on an entirely different platform, with different physical poses, surrounding environments, and travel speeds.  

\section{Dual-Mobility Takeover Modeling}
In total, we had 142321 samples, with 55791 samples from micro-mobility and 86530 samples from cars. The car had more samples because the videos were longer. We used the car samples as training and the micro-mobility samples as test. In this way, we build a driver profile for newer mobilities, with data from the commonly used mobility in the current society. In our training data, there are 5930 samples with takeovers, which is 10.1\% of the training size. The dataset is imbalanced but the takeover percentage is already much higher than many existing driving simulator studies \cite{qiu2022incorporating}. Takeover in AV is a naturally rare case because too many takeovers would lead to rejection of such AV or ADAS features. We speculate that the more immersive simulation experiences and high frequency of interactions with other road users may have led to a higher percentage of takeovers. Interestingly, we found out that the takeover rate is 11.81\% in aggressive driving mode and 8.55\% in defensive driving mode. Meanwhile, the participants had a 7.50\% takeover rate with the proactive audio feedback, and it increased to 11.50\% without the proactive audio feedback. The more defensive driving style and proactive audio feedback may have increased participants' trust and thus resulted in fewer takeovers. To address the imbalance problem of our training data, we down-sampled the major class to make a perfectly balanced training set \cite{mollahosseini2017affectnet}.

\subsection{Random Forest Baseline}
In takeover prediction literature, Random Forest outperforms other models in many cases \cite{du2020predicting, yu2021measurement}, potentially because of its efficiency on smaller datasets, and the ability to adapt to multimodal signals. We use the RF as a baseline model for this binary prediction task. We used the Scikit-learn library \cite{kramer2016scikit} to build the classifier, and ran a grid search on parameters to improve the performance. The resulting best-performing RF classifier had an accuracy of 0.787 and an AUC value of 0.595.

\subsection{Deep Neural Network}
Dual-take is a feed-forward DNN, built with the Tensorflow Keras library \cite{chollet2015keras}. The network structure starts with an input layer length of 54 to match the input feature counts. There are three hidden layers with 64, 32, and 16 ReLu units. Each layer receives the input values from the prior layer and outputs to the next one. Then, we put a 1D max-pooling layer to reduce the spatial size and reduce over-fitting, followed by a dropout layer with a rate of 0.1. The network utilizes a binary cross entropy and an Adam optimizer with a learning rate of 0.001. Dual-take is trained for 20 epochs with a batch size of 16, using a mini-batch stochastic gradient descent. The Dual-take DNN model had an accuracy of 0.856 and an AUC value of 0.741. The main future application is utilizing car data to predict behaviors in newer mobilities.

\subsection{Transfer Learning}
\begin{figure}[!tbp]
  \centering
  \begin{minipage}[b]{0.27\textwidth}
    \includegraphics[width=\textwidth]{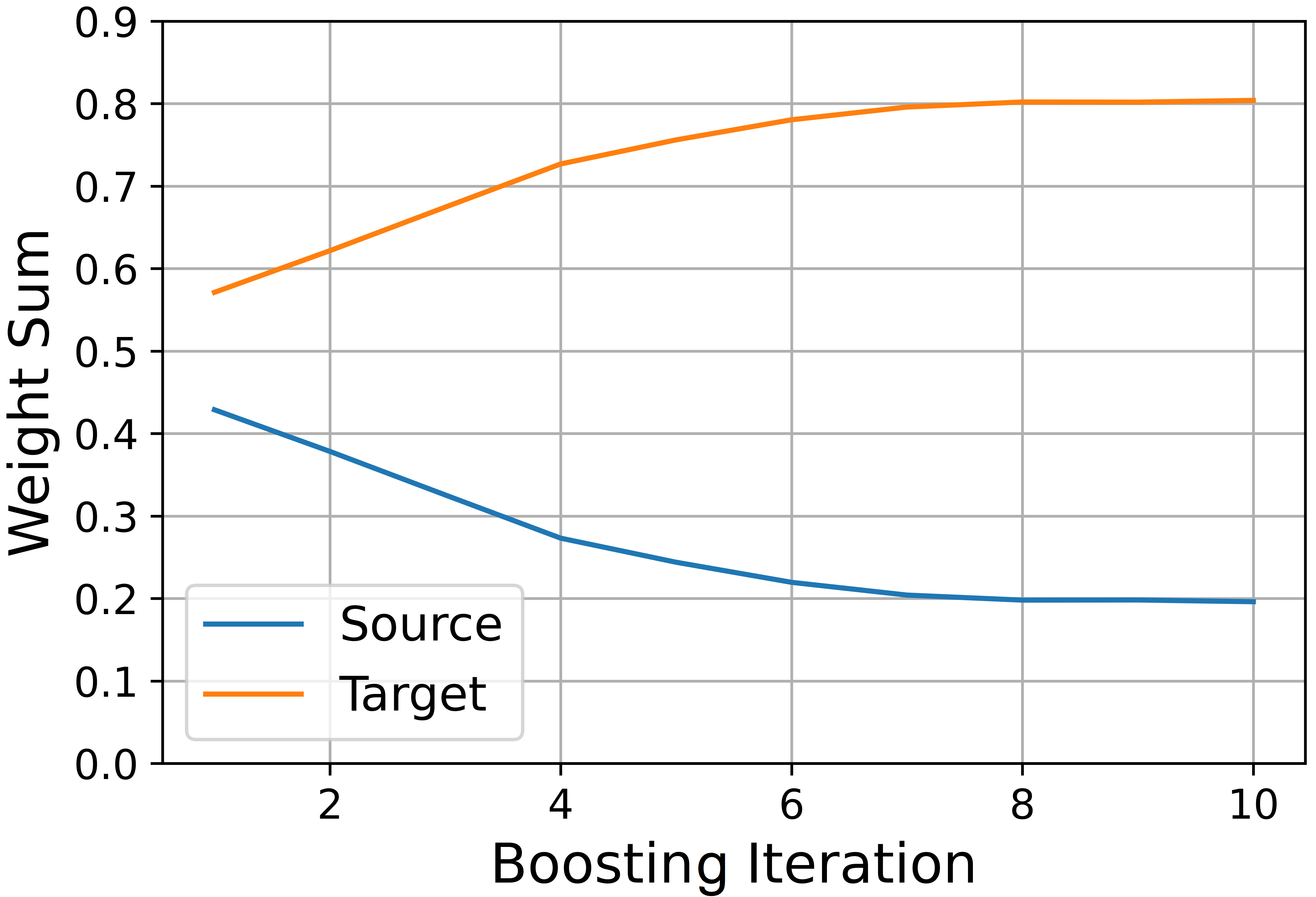}
    \caption{Target and Source Weight Sum over Boosting Iteration}
    \label{fig:TargetWeight}
  \end{minipage}\vspace{-0.5em}
\end{figure}

From the statistical analysis in section 3, we can see that drivers behave differently across mobilities. To better model behaviors on different mobilities, we integrated transfer learning in DualTake. We utilized the TrAdaBoost algorithm, which is a supervised instances-based domain adaption method \cite{de2021adapt}. The algorithm has weights for the source and target samples and fits an estimator on source and target labeled data. It computes the error vectors of training instances. With the error, the algorithm computes the total weighted error of target instances and finally updates the source and target weights. The algorithm will return to step 1 and loop until the number of boosting iteration is reached. We used a boost iteration of 10 with a learning rate of 0.5 for TrAdaBoost. We conducted a group-k-fold validation, that we used the car samples for the training source data. We split the micro-mobility data into 5 folds by participants, and each fold contains certain participants. We then used 4 folds of micro-mobility samples as training target data, with the rest 1 fold as test data. In this cross-validation method, we simulate the future application scenario, that we pre-collect micro-mobility and car behavioral data from a certain amount of participants, and we want to predict takeovers from participants with only car driving data. The model reached an accuracy of 0.8612 and an AUV value of 0.767. From the 5 DualTake TrAdaBoost models in the cross-validation, we computed the source and target weights for each iteration, and summed up the weights for all the samples. We visualize the weight on source and target data in Fig. \ref{fig:TargetWeight}. The TrAdaBoost algorithm gradually shifts weight toward target samples, and it converges around 7 iterations, where the accuracy and AUC values also converge. This result may indicate that target samples contain more important information about takeover predictions. The source data also contributes to the prediction, potentially because the source data contain the behavioral responses from the same participant.

\begin{figure}[!tbp]
  \centering
  \begin{minipage}[b]{0.27\textwidth}
    \includegraphics[width=\textwidth]{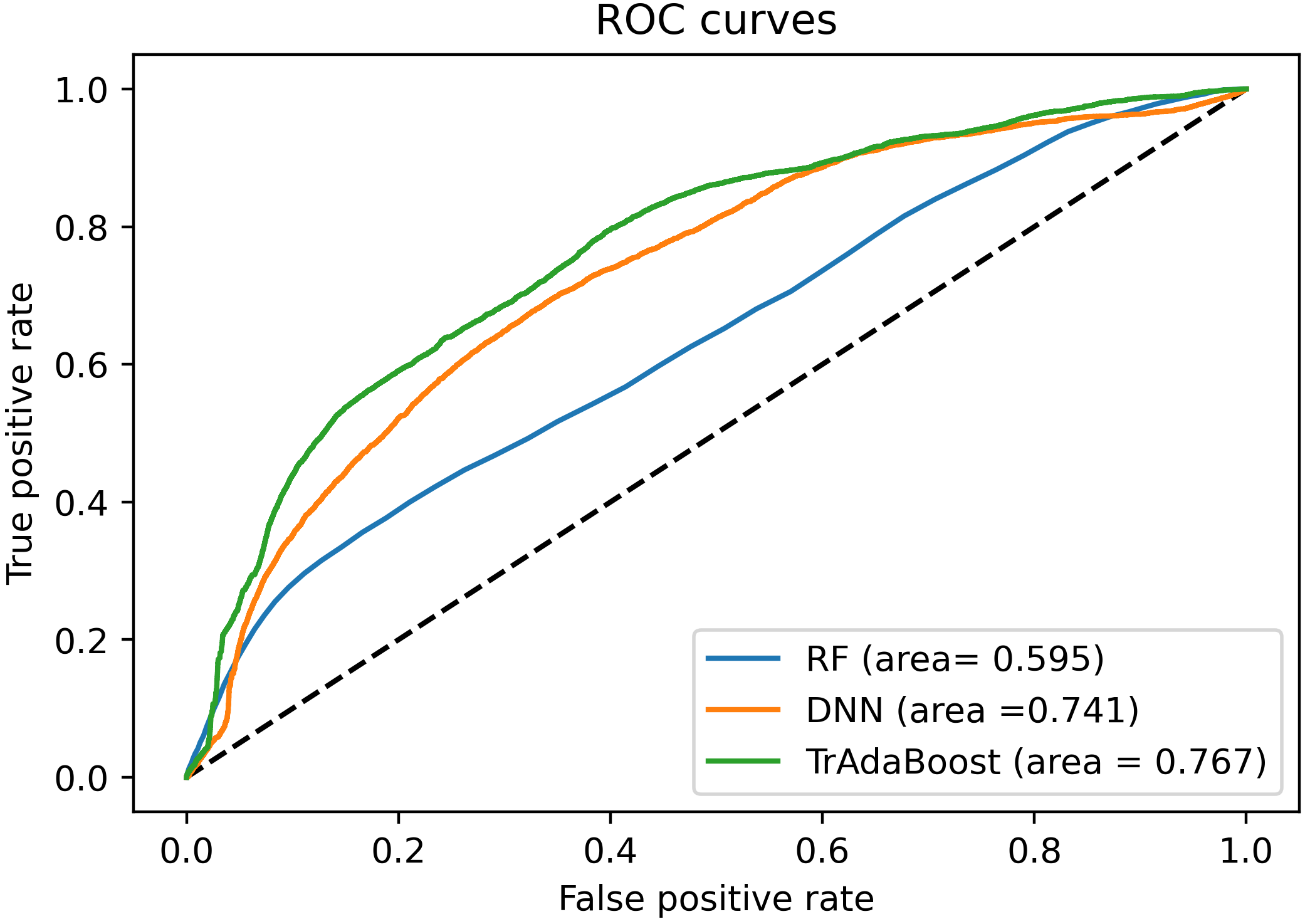}
    \caption{ROC Curves of Models}
    \label{fig:ROC}
  \end{minipage}\vspace{-0.5em}
\end{figure}

ROC curves from the baseline RF model, the DualTake DNN only, and the DualTake with TrAdaBoost in Fig. \ref{fig:ROC}. The DualTake with TrAdaBoost reached the best performances. We also list the performance of our models with some existing work in takeover prediction, in Table \ref{tab:comparison}. The takeover prediction is a temporal task, and many researchers in the field have tried time-series modeling approach. We have investigated a Long Short Term Memory (LSTM) model with time series data in our study. We have included their performance in the algorithm comparison table. The time-series analysis did not out-perform the DNN model, potentially due to the nature of gaze and semantics data, that are noisy and not entirely linear. The sample size we have compared to our data dimension is also very limited. Similar limitations and results were observed in existing literature \cite{deo2019looking}. DualTake predicts takeovers across mobilities, and it is showing comparable performance to most related existing work on single mobility. 

\begin{table}\footnotesize{
  \caption{Algorithm Comparison}\vspace{-1em}
  \label{tab:comparison}
  \begin{tabular}{ccc}
    \toprule
    Model&Accuracy&AUC\\
    \midrule
    RF baseline&0.787&0.595 \\
    DualTake DNN&0.856&0.741\\
    DualTake TrAdaBoost&\textbf{0.8612}&\textbf{0.767}\\
    \midrule
    LSTM&0.738&0.541\\
    \midrule
    Du et al.&0.85&0.69\\
    Pakdamanian et al.&\textbf{0.96}&\textbf{0.96}\\
  \bottomrule\vspace{-2em}
\end{tabular}}
\end{table}

\section{Discussion and Conclusion}

This study developed a DNN and transfer learning based model, DualTake, to predict driver's takeover across mobilities. In contrast of existing research that focus on takeover prediction only in cars, we explored using behavioral data in traditional mobility to predict takeovers in newer mobility. Despite the significant driver behavioral differences across mobilities, the DualTake model reached an accuracy of 0.86 and an AUC value of 0.76. Through transfer learning, we also found out the DualTake utilizes general behavioral patterns from the newer mobility, and also the individualized patterns. The promising performance of our model demonstrates the application feasibility of personalized driver monitoring in future hybrid society. Users will be able to travel on different types of AVs, and have a universal model to monitor and predict their behaviors. 

There are some limitations to this work. First, the work is a proof of concept for takeover prediction across mobilities and there are only two types of mobilities involved. In the future work, more types of mobilities should be involved such as urban air mobilities. The other limitation is that these behavioral responses have time casual relationships and a time series modeling method may further improve the performance. With more data and time, we could also apply attention networks to such prediction tasks. Despite the limitations, this work is the first in the field to demonstrate the feasibility of driver state understanding across mobilities, with promising performance.

\section{Acknowledgments}
The authors would like to thank Jacob Hunter, Elise Ulwelling, Matthew Konishi, Noah Michelini, Akhil Modali, Anne Mendoza, Jessie Snyder, Shashank Mehrotra, Anil Kumar, Neera Jain, and Tahira Reid for their data collection efforts of the dataset used in this work. 

\newpage

\footnotesize
\bibliographystyle{ACM-Reference-Format}
\bibliography{refs}
\balance

\end{document}